\documentstyle[aps]{revtex}

\begin{document}
\title{Fermionic Bound States and Pseudoscalar Exchange}
\author{A. Anselm\thanks{%
died Boston, Massachusetts 23 August 1998}}
\address{Theoretical Division, Petersburg Nuclear Physics Institute, Gachina, Russia}
\author{N. Dombey}
\address{Centre for Theoretical Physics, University of Sussex, Brighton, BN1 9QJ, UK}
\maketitle

\begin{abstract}
\noindent We discuss the possibility that fermions bind due to Higgs or
pseudoscalar exchange. It is reasonable to believe on qualitative grounds
that this can occur for fermions with a mass larger than 800-900 GeV. An
exchange of a pseudoscalar boson leads in the nonrelativisitc limit to an
unacceptable potential which behaves like $1/r^3$ at the origin.\ We show
that this singular behaviour is smeared out when relativistic effects are
included.
\end{abstract}

\section{Introduction}

\noindent In this note, we obtain an estimate for the mass of a fermion
which is heavy enough to bind to its anti-particle via Higgs exchange. We
then try to repeat the argument for pseudoscalar exchange, since
pseudoscalars arise naturally in some extensions of the standard model. We
will find eventually that the situation is similar to that in Higgs
exchange, but the analysis is now much more complex as the exchange of a
non-relativistic pseudoscalar seems to lead to a potential which is singular
at the origin. We shall show that if the pseudoscalar exchange is treated in
a relativistic framework this singularity vanishes.

\smallskip\ 

\noindent We begin by considering Higgs exchange between the usual quarks
taken to be of mass $m$. This leads to a weak attraction with coupling
constant (see for example Reference \cite{ref1}): 
\begin{equation}
g=\sqrt{2}\left( \frac mv\right) <<1;\,\,\,\,v=\left( G_F\sqrt{2}\right)
^{-1/2}=245\text{ GeV}  \label{est1}
\end{equation}
Even for the t-quark with $m_t\simeq 180$ GeV$,\,\,g\approx 1$. (This does
not necessarily invalidate a perturbative approach since the typical
parameter of the perturbative expansion is $g^2/8\pi ^2\ll 1$). Even this
coupling, however, is not strong enough to create bound states of $t\bar t$
or $tt$. In principle, however, heavier fermions $f$ for which the coupling $%
g$ is greater than one can be found where bound states $f\overline{f}$ are
bound by Higgs exchange.

\smallskip\ 

\noindent At a qualitative level it is easy to estimate how heavy fermions
should be for them to be bound by means of Higgs exchange. Obviously for
non-relativistic particles, the kinetic energy $p^2/m$ (we use the reduced
mass $m/2$ of the two particles with mass $m$) should be smaller than the
potential energy which is 
\begin{equation}
\left| S\right| =\frac{g^2}{4\pi }\frac 1re^{-m_Hr}  \label{est2}
\end{equation}
where $m_H$ is the Higgs mass. If the composite state has the radius $a$,
then $p^2/m\sim 1/ma^2$ and we get the condition: 
\begin{equation}
\frac 1{ma^2}<\frac{g^2}{4\pi }\frac 1ae^{-m_Ha}  \label{est3}
\end{equation}
This equation can be rewritten in the form 
\begin{equation}
\frac{g^2m}{m_H}>\frac{4\pi e^{m_Ha}}{m_Ha}  \label{est4}
\end{equation}
Whatever the value of $a$ the right hand side of Eq (\ref{est4}) is always
larger than $4\pi e$. Therefore 
\begin{equation}
\frac{g^2m}{m_H}>4\pi e=34.3^{}  \label{est4a}
\end{equation}
We use Eq (\ref{est1}) to estimate $g.$ So 
\begin{equation}
\frac{m^3}{v^2m_H}>17  \label{est5}
\end{equation}
For $m=m_t=180$ GeV this leads to $m_H<5.7$ GeV, which is unacceptable.
However if we take $m_H=400$ GeV (for example), Eq. (\ref{est5}) gives: 
\begin{equation}
m>800\text{ GeV}  \label{est6}
\end{equation}
This would be a very heavy fermion, but fermions of mass in the TeV region
arise in many extensions of the standard model.

\smallskip\ 

\noindent We will now try to extend this simple analysis to pseudoscalar
exchange. Though in the SM there is no pseudoscalar Higgs they occur in many
extensions of the SM, for example in supersymmetric extensions. We show in
the next Section that the exchange of a pseudoscalar meson $A$ leads in the
nonrelativistic approximation to the potential 
\begin{eqnarray}
P(r) &=&-\frac 1{4\pi v_A^2}e^{-m_Ar}\left[ \frac 1{r^3}\left( \vec \sigma
_1.\vec \sigma _2-3(\vec \sigma _1.\vec n)(\vec \sigma _2.\vec n)\right)
\left( 1+m_Ar\right) \right.  \label{psex} \\
&&\ \ \ \left. -\frac{m_A^2}r(\vec \sigma _1.\vec n)(\vec \sigma _2.\vec n%
)\right] ,  \nonumber
\end{eqnarray}
where $\vec \sigma _1,\vec \sigma _2$ are the Pauli matrices for the two
fermions, $\vec n=\vec r\,/\,r$ and $m_A,\,v_A$ are the mass of the A-boson
and the scale, analogous to $v=246$ GeV in the standard model. The potential
of Eq (\ref{psex}) depends on the spins of the fermions and in certain spin
states can lead to an attraction between fermions. In this case the analogue
of the inequality (\ref{est3}) is 
\begin{equation}
\frac 1{ma^2}<\frac 1{4\pi v_A^2}\frac 1{a^3}e^{-m_Aa}  \label{psex2}
\end{equation}
Clearly for a small enough radius $a$ this condition can be satisfied. But
this corresponds to the well-known fact that an attractive potential like $%
1/r^3$ which is more singular at the origin than the centrifugal term leads
to the ``collapse'' of the particle to the origin and hence to a problem
which is not well-defined \cite{case}. We shall demonstrate, however, that
while the expression (\ref{psex}) holds for nonrelativistic fermions and
leads to collapse, a relativistic treatment will lessen the singularity at $%
r=0$ so that there is no collapse to the centre.

\smallskip\ 

\noindent Suppose now that 
\begin{equation}
m_Aa\ll 1  \label{ma}
\end{equation}
Then (\ref{psex2}) means that 
\begin{equation}
a<\frac m{4\pi v_A^2}  \label{psex3}
\end{equation}
The condition that the fermions should be nonrelativistic $p^2/m<m$ at $%
p\simeq 1/a$ together with eq. (\ref{psex3}) can be written in the form: 
\begin{equation}
\frac 1m<a<\frac m{4\pi v_A^2}  \label{a}
\end{equation}
It is thus required that 
\begin{equation}
m>\sqrt{4\pi }v_A
\end{equation}
For $v_A\sim v=245$ GeV this leads to 
\begin{equation}
m>870\text{ GeV}
\end{equation}
This restriction is not very different from (\ref{est6}) for the scalar
case. Note that for $a\sim (870$ GeV$)^{-1}$ the condition (\ref{ma}) is
probably satisfied since we can expect the pseudoscalar $A$ to be lighter
than the Higgs (the present limit is only $m_A>24.3$ GeV) \cite{data}$.$

\smallskip\ 

\noindent Note an interesting feature of bound states mediated by scalar and
pseudoscalar exchange: unlike the situation with vector exchange, states of
the same mass can appear for both $f\bar f$ and $ff$ systems. This is a
consequence of the fact that scalar and pseudoscalar interactions are even
under charge conjugation.

\section{Pseudoscalar exchange (non-relativistic case)}

\noindent We now look in more detail at pseudoscalar exchange. Start with
the interaction of the pseudo-scalar boson $A$ with a fermion via the
coupling $i(2m/v_A)\left( \bar \psi \gamma _5\psi \right) A$. The Feynman
amplitude corresponding to the exchange of $A$ can then be represented in
the form 
\begin{equation}
M=\frac{4m^2}{v_A^2}\left( \varphi _2^{\prime *}\vec \sigma \vec q\varphi
_2\right) \left( \varphi _1^{\prime *}\vec \sigma \vec q\varphi _1\right) 
\frac 1{m_A^2+\vec q^2},
\end{equation}
\noindent where $\vec q=\vec p_1^{\prime }-\vec p_1=\vec p_2-\vec p%
_2^{\prime }$ is the three-momentum transfer and we have taken the
non-relativistic limit so that the four-component spinors $\psi $ are
replaced by the two-component spinors $\varphi .$

\smallskip\ 

\noindent The Born amplitude $A\left( \vec q\right) $ is related to $M\left( 
\vec q\right) $ by 
\begin{equation}
A=\frac 1{8\pi W}M
\end{equation}
\noindent where $W$ is the total centre of mass energy, and the potential $%
P\left( \vec r\right) $ is the Fourier transform of $A\left( \vec q\right) $%
: 
\begin{equation}
P\left( \vec r\right) =-\frac{4\pi }m\int e^{i\vec q\vec r}A\left( \vec q%
\right) \frac{d^3q}{\left( 2\pi \right) ^3}
\end{equation}
\noindent From the last three equations we can easily calculate $P\left( 
\vec r\right) $ (we use the nonrelativistic value $W\simeq 2m$): 
\begin{eqnarray}
P\left( \vec r\right) &=&\frac 1{v_A^2}\left( \varphi _2^{\prime *}\sigma
_i\varphi _2\right) \left( \sigma _1^{\prime *}\sigma _n\varphi _1\right) 
\frac \partial {\partial x_i}\frac \partial {\partial x_n}\left[ \frac 1{%
4\pi r}e^{-m_Ar}\right] =  \label{pot} \\
&=&-\frac 1{4\pi v_A^2}e^{-m_Ar}\left[ \frac 1{r^3}\left( \left( \vec \sigma
_1.\vec \sigma _2\right) -3\left( \vec \sigma _1.\vec n\right) \left( \vec %
\sigma _2.\vec n\right) \right) \left( 1+m_Ar\right) \right.  \nonumber \\
&&\left. -\frac{m_A^2}r\left( \vec \sigma _1.\vec n\right) \left( \vec \sigma
_2.\vec n\right) \right]  \nonumber
\end{eqnarray}
\noindent In the last formulae we have omitted the spinors $\varphi _{1\text{%
,}}\varphi _1^{\prime }...$ but have labelled the $\sigma $'s by the indices
showing to which fermion they refer

\smallskip\ .

\noindent Following the discussion of the Introduction we shall consider
only fermions which are heavy enough to bind. The radius of the bound states
is of order $a$ which must be small enough to ensure the validity of the
inequality. 
\begin{equation}
m_Aa\ll 1
\end{equation}
\noindent So the potential (\ref{pot}) would be of the form: 
\begin{equation}
P\left( \vec r\right) =-\frac 1{4\pi v_A^2}\frac{e^{-mr}}{r^3}\left[ \vec %
\sigma _1.\vec \sigma _2-3\left( \vec \sigma _1.\vec n\right) \left( \vec %
\sigma _2.\vec n\right) \right] .
\end{equation}
\noindent To compare our nonrelativistic approach with relativistic
equations we shall use the technique developed by Kr\'olikowski in handling
the spin dependence of the wave function \cite{krol} For the Schr\"odinger
equation 
\begin{equation}
-\frac 1m\nabla ^2\psi +P\left( r\right) \psi =E\psi  \label{schr}
\end{equation}
\noindent we decompose $\psi $ into spin-zero and spin-one parts: 
\begin{equation}
\phi =P_0\psi ,\,\,\,\,\,\vec \chi =\frac 12\left( \vec \sigma _1-\vec \sigma
_2\right) P_1\psi ,  \label{decomp}
\end{equation}
where the projection operators are: 
\begin{equation}
P_0=\frac 14\left( 1-\vec \sigma _1.\vec \sigma _2\right) ,\;\;P_1=\frac 14%
(3+\vec \sigma _1.\vec \sigma _2)
\end{equation}
The following set of identities can be used: 
\[
P_0\left( \vec \sigma _1-\vec \sigma _2\right) =\left( \vec \sigma _1-\vec %
\sigma _2\right) P_1,\;\;P_0\left( \vec \sigma _1+\vec \sigma _2\right)
=\left( \vec \sigma _1+\vec \sigma _2\right) P_0=0, 
\]

\[
P_0\left( \sigma _{1i}.\sigma _{2n}-\sigma _{1n}\sigma _{2i}\right)
=iP_0\epsilon _{in\ell }\left( \sigma _{1\ell }-\sigma _{2\ell }\right) 
\]
\[
P_0\left( \sigma _{i1}.\sigma _{2n}+\sigma _{1n}\sigma _{2i}\right)
=-2\delta _{in}P_0, 
\]
\[
P_0\left( \sigma _{1i}-\sigma _{2i}\right) \left( \sigma _{1n}+\sigma
_{2n}\right) =2iP_0\epsilon _{in\ell }\left( \sigma _{1\ell }-\sigma _{2\ell
}\right) 
\]
\[
\stackrel{\rightarrow }{P}_0\left( \sigma _{1i}-\sigma _{2i}\right) \left(
\sigma _{1n}-\sigma _{2n}\right) =4P_0\delta _{in} 
\]

\begin{eqnarray*}
&&P_0\left( \sigma _{1i}-\sigma _{2i}\right) \left( \sigma _{1n}\sigma
_{2\ell }+\sigma _{1\ell }\sigma _{2n}\right) \\
&=&2\left[ \delta _{n\ell }\left( \sigma _{1i}-\sigma _{2i}\right) -\delta
_{ni}\left( \sigma _{1\ell }-\sigma _{2\ell }\right) -\delta _{i\ell }\left(
\sigma _{1n}-\sigma _{2n}\right) \right] P_1
\end{eqnarray*}

\noindent Acting by $P_0$ on both sides of eq. (\ref{schr}) one sees that
the tensor forces vanish in the singlet state, so that $\phi $ obeys the
free Schr\"odinger equation. On the other hand by applying the operator $%
\left( \vec \sigma _1-\vec \sigma _2\right) P_1=P_0\left( \vec \sigma _1-%
\vec \sigma _2\right) $ to eq. (\ref{schr}) we can derive the equation for
the triplet $\vec \chi $ function: 
\begin{equation}
-\frac 1m\nabla ^2\vec \chi =\frac 1{2\pi v_A^2}\frac 1{r^3}\left( \vec \chi
-3\vec n\left( \vec \chi .\vec n\right) \right) =E\vec \chi  \label{sch2}
\end{equation}
\noindent Following \cite{krol} we now split $\vec \chi \left( \vec r\right) 
$ into ``electric'', ``longitudinal'' and ``magnetic'' parts: 
\begin{equation}
\vec \chi \left( \vec r\right) =\vec \chi _e\left( \vec r\right) +\vec \chi
_L\left( \vec r\right) +\vec \chi _M\left( \vec r\right) ,  \label{khi}
\end{equation}
\noindent where each obeys the following expansion in spherical harmonics: 
\begin{eqnarray}
\vec \chi _e\left( \vec r\right) &=&\sum_{jm}\vec nY_{jm}\left( \vec n%
_{\perp }\right) \chi _e^{jm}\left( r\right) ,  \label{kr} \\
\vec \chi _L\left( \vec r\right) &=&-\sum_{jm}\frac \partial {\partial \vec n%
_{\perp }}\left( \frac{Y_{jm}\left( \vec n_{\perp }\right) }{j\left(
j+1\right) }\right) \chi _L^{jm}\left( r\right) ,  \nonumber \\
\vec \chi _m\left( \vec r\right) &=&-\sum_{jm}\left( \vec n\times \frac %
\partial {\partial \vec n_{\perp }}\right) \frac{Y_{jm}\left( \vec n%
_L\right) }{j\left( j+1\right) }\chi _M^{jm}\left( r\right)  \nonumber
\end{eqnarray}

\noindent In eqns. (\ref{kr}) we introduce differention in the transverse
direction 
\begin{equation}
\frac \partial {\partial n_{i\perp }}\equiv \left( \delta
_{in}-n_in_j\right) \frac \partial {\partial n_j},
\end{equation}
\noindent so that 
\begin{equation}
n_i\frac \partial {\partial n_{\perp i}}=0.
\end{equation}
\noindent On the other hand the gradient and the Laplacian are expressed in
terms of $\partial /\partial r,\,\,\partial /\partial \vec n_{\perp }$ as
follows: 
\begin{eqnarray}
\frac \partial {\partial \vec r} &=&\frac 1r\frac \partial {\partial \vec n%
_{\perp }}+\vec n\frac \partial {\partial r}, \\
\nabla ^2 &=&\frac{\partial ^2}{\partial \vec r^2}=\frac{\partial ^2}{%
\partial r^2}+\frac 2r\frac \partial {\partial r}+\frac 1{r^2}\frac{\partial
^2}{\partial \vec n_{\perp }^2}.  \nonumber
\end{eqnarray}
\noindent Note that $\partial ^2/\partial \vec n_{\perp }^2$ represents the
angular part of the Laplacian. Note also that: 
\begin{equation}
\frac \partial {\partial \vec n_{\perp }}.\vec n=2,\,\,\frac \partial {%
\partial \vec n_{\perp }}=\vec n_{\perp }\left( \vec n_{\perp }.\frac %
\partial {\partial \vec n}\right) ,\,\frac{\partial ^2}{\partial \vec n%
_{\perp }^2}Y_{jm}=-j\left( j+1\right) Y_{jm}
\end{equation}

\noindent The quantum number $j$ corresponds to the total rather than the
orbital angular momentum. Indeed let us define the total angular momentum $%
\vec J=\vec L+\vec S$ acting on the vector $\vec \chi $ of eq. (\ref{khi})
as: 
\begin{equation}
\left( L\chi _k\right) _i=-i\epsilon _{ipq}n_p\frac{\partial \chi _k}{%
\partial nq},\,\,\,\,\left( S\chi _k\right) _i=-i\epsilon _{ik\ell }\chi _l.
\end{equation}

\noindent It is easy to see then that the operator $\vec L$ when acting on $%
\vec \chi _e,\,\,\vec \chi _L$ and $\vec \chi _M$ can be commuted through
the vectors $\vec n,\,\,\,\,\partial /\partial \vec n_{\perp }$ and $\left( 
\vec n\times \partial /\partial \vec n_{\perp }\right) $ and that the
contribution of $\vec S$ in $\vec J=\vec L+\vec S$ cancels out so that $\vec %
J$ acts exactly in the same way as $\vec L$ on $Y_{jm}$. For example, taking 
$\left( \vec \chi _e\left( \vec r\right) \right) _k\equiv \chi _{ek}\left( 
\vec r\right) $, we get 
\begin{eqnarray}
\left( J\chi _{ek}\right) _i &=&\sum_{jm}\chi _e^{jm}\left( r\right) \left\{
-i\epsilon _{ipq}n_p\frac \partial {\partial n_q}\left( n_kY_{jm}\right)
-i\epsilon _{ik\ell }n_lY_{jm}\right\} = \\
&=&\sum_{jm}\chi _e^{jm}\left( r\right) \left[ -i\epsilon _{ipq}n_pn_k\frac{%
\partial Y_{jm}}{\partial n_q}\right] =\sum_{jm}n_k\chi _e^{jm}\left(
r\right) \left( L_iY_{jm}\right)  \nonumber
\end{eqnarray}

\noindent If we now repeat this operation: 
\begin{equation}
{\cal J}^2\chi _{ek}={\cal J}_i\left( {\cal J}\chi _{ek}\right)
_i=\sum_{jm}n_k\chi _e^{jm}\left( r\right) \vec L^2Y_{jm}=j\left( j+1\right)
\chi _{ek},
\end{equation}
\noindent which proves that $j$ is actually the total momentum. The same
proof can be given for $\vec \chi _L$ and $\vec \chi _M$. We are ready now
to derive the equations for the radial wave functions. For that purpose we
project eq. (\ref{sch2}) on $\vec n$ and act on both sides of the equation
by $\partial /\partial \vec n_{\perp }$ and $\vec n\times 
{\displaystyle {\partial \over \partial \vec n_{\perp }}}%
$. Using the expansions (\ref{khi}) and (\ref{kr}) it is easy to obtain
after some algebra the following equations: 
\begin{equation}
\begin{array}[t]{l}
{\displaystyle {d^2\chi _e \over dr^2}}%
+%
{\displaystyle {2 \over r}}%
{\displaystyle {d\chi _e \over dr}}%
-%
{\displaystyle {j\left( j+1\right) +2 \over r^2}}%
\chi _e-%
{\displaystyle {2\chi _L \over r^2}}%
+%
{\displaystyle {m \over \pi v_A^2r^3}}%
\chi _e+mE\chi _e=0, \\ 
{\displaystyle {d^2\chi _L \over dr^2}}%
+%
{\displaystyle {2 \over r}}%
{\displaystyle {d\chi _L \over dr}}%
-%
{\displaystyle {j\left( j+1\right)  \over r^2}}%
\chi _L-%
{\displaystyle {2j\left( j+1\right)  \over r^2}}%
\chi _e-%
{\displaystyle {m\chi _L \over 2\pi v_A^2r^3}}%
+mE\chi _L=0, \\ 
{\displaystyle {d^2\chi _M \over dr^2}}%
+%
{\displaystyle {2 \over r}}%
{\displaystyle {d\chi _M \over dr}}%
-%
{\displaystyle {j\left( j+1\right)  \over r^2}}%
\chi _M-%
{\displaystyle {m \over 2\pi v_A^2}}%
\chi _M+mE\chi _M=0.
\end{array}
\label{khi3}
\end{equation}
\noindent where we have omitted the indices $jm$ of the functions $\chi
_e^{jm}$, ... in these equations.

\smallskip\ 

\noindent We see that the functions $\chi _e$ and $\chi _L$ are coupled in
the first two equations (\ref{khi3}) whereas the function $\chi _M$ is
separated from $\chi _e,\chi _L$. This is related to the fact that $\chi _M$
corresponds to the value of the orbital momentum $\ell =j$ while $\chi _e$
and $\chi _L$ are the mixtures of the states $\chi ^{+}$ with $\ell =j+1$
and $\chi ^{-}$ with $\ell =j-1$. These mixtures are \cite{krol}: 
\begin{eqnarray}
\chi _e &=&\sqrt{\frac j{2j+1}}\chi ^{-}+\sqrt{\frac{j+1}{2j+1}}\chi ^{+}, \\
\chi _L &=&\sqrt{j(j+1)}\left[ -\sqrt{\frac{j+1}{2j+1}}\chi ^{-}+\sqrt{\frac %
j{2j+1}}\chi ^{+}\right]  \nonumber
\end{eqnarray}
\noindent Passing in eqns. (\ref{khi3}) from $\chi _e$ and $\chi _L$ to $%
\chi ^{\pm }$ we obtain: 
\begin{equation}
\begin{array}[t]{l}
{\displaystyle {d^2\chi ^{-} \over dr^2}}%
+%
{\displaystyle {2 \over r}}%
{\displaystyle {d\chi ^{-} \over dr}}%
-%
{\displaystyle {j\left( j-1\right)  \over r^2}}%
\chi ^{-}+mE\chi ^{-}+%
{\displaystyle {m \over 2\pi v_A^2r^3}}%
\left[ 
{\displaystyle {j-1 \over 2j+1}}%
\chi ^{-}+%
{\displaystyle {3\sqrt{j(j+1)} \over 2j+1}}%
\chi ^{+}\right] =0, \\ 
{\displaystyle {d^2\chi ^{+} \over dr^2}}%
+%
{\displaystyle {2 \over r}}%
{\displaystyle {d\chi ^{+} \over dr}}%
-%
{\displaystyle {\left( j+1\right) \left( j+2\right)  \over r^2}}%
\chi ^{+}+mE\chi ^{+}+%
{\displaystyle {m \over 2\pi v_A^2r^3}}%
\left[ 
{\displaystyle {j+2 \over 2j+1}}%
\chi ^{+}+%
{\displaystyle {3\sqrt{j(j+1)} \over 2j+1}}%
\chi ^{-}\right] =0.
\end{array}
\label{khi4}
\end{equation}
\noindent We see from (\ref{khi4}) that $j\left( j-1\right) $ in the first
equation and $\left( j+1\right) \left( j+2\right) $ in the second correspond
to $\ell \left( \ell +1\right) $ for $\ell =j\mp 1$. Note that in each part
of Eq. (\ref{khi4}) the potentials are attractive and singular as $\sim
1/r^3 $ at $r\rightarrow 0$. Note also that $\chi _M$ in eqn (\ref{khi3})
does not have this singularity. We show in the next Section that this
singularity is not present at all in a fully-relativistic treatment of
pseudoscalar exchange: in the Appendix we show that the apparent singularity
arises only when a non-relativistic approximation is made. So Eq. (\ref{khi4}%
) is valid provided that $r$ is not too small; more precisely that $r>a$,
where the cutoff $a$ is given by Eq. (\ref{a}).

\section{The relativistic equations}

\noindent To obtain a relativistic treatment of the problem and in
particular, to understand the smoothing out of the singularity $1/r^3$ we
begin with the two-body Dirac equation (the Breit equation) 
\begin{equation}
\left[ E-\gamma _0^{(1)}\left( \vec \gamma ^{(1)}\vec p+m_1\right) -\gamma
_0^{(2)}\left( -\vec \gamma ^{(2)}\vec p+m_2\right) -V_{int}\left( r\right)
\right] \psi =0  \label{breit}
\end{equation}
\noindent describing a system of two spin-1/2 particles of masses $m_1$ and $%
m_2$ in the centre of mass frame, interacting with each other through a
potential of the form 
\begin{equation}
\begin{array}{ll}
V_{int} & =V_s\left( r\right) +V_p\left( r\right) +V_v\left( r\right) \\ 
V_s & =\left( \gamma _0^{\left( 1\right) }\times \gamma _0^{\left( 2\right)
}\right) S\left( r\right) \\ 
V_p & =\left[ \left( \gamma _0^{\left( 1\right) }\gamma _5^{\left( 1\right)
}\right) \times \left( \gamma _0^{\left( 2\right) }\gamma _5^{\left(
2\right) }\right) \right] P\left( r\right) \\ 
V_v & \left[ \left( \gamma _0^{\left( 1\right) }\gamma _\mu ^{\left(
1\right) }\right) \times \left( \gamma _0^{\left( 2\right) }\gamma _\mu
^{\left( 2\right) }\right) \right] V\left( r\right)
\end{array}
\label{pote}
\end{equation}

\noindent where $S\left( r\right) $,$P\left( r\right) $, and $V\left(
r\right) $ result from the exchange of scalar, pseudoscalar and vector
particles. We use the Dirac-Pauli representation for the $\gamma $-matrices:

\begin{equation}
\gamma _0=\beta =\left( 
\begin{array}{cc}
1 & 0 \\ 
0 & -1
\end{array}
\right) ,\quad \vec \gamma =\left( 
\begin{array}{cc}
0 & \vec \sigma \\ 
-\vec \sigma & 0
\end{array}
\right) ,\quad \gamma _5=i\gamma _0\gamma _1\gamma _2\gamma _3
\end{equation}

\noindent In this paper we shall be interested mainly in $S$ and $P$
exchange. Of course, a static potential $V_{int}$ which depends only on the
variable $r=\left| \vec r_1-\vec r_2\right| $ is not a relativistically
invariant quantity. Furthermore the concept of a potential can only be an
approximation at best to a proper field-theoretic description of a system.
These are well-known difficulties of the generalisation of the Dirac
equation involving a potential to of a two-body system which we do not wish
to pursue here. We shall adopt the viewpoint that Eq. (\ref{breit}) provides
a starting point for a calculation of the relativistic corrections to the
non-relativistic two-fermion system in the same way that the treatment of
the Coulomb potential in the Dirac equation leads to relativistic
corrections in the spectrum of the hydrogen atom of higher order in $\alpha $
compared with the Schrodinger solution. An analysis of the constraints
imposed by the relativistic invariance of a two-body fermionic system shows
that provided the coordinates are chosen appropriately (which in practice
means using the centre of mass frame for the equal mass case of $m_1=m_2$),
then a potential $V_{int}\left( r\right) $ can be used in a relativistic
two-body equation without violating relativistic invariance. Further details
can be found in the papers by Mourad and Sazdjian \cite{mour} and Crater and
Long \cite{xxx}, which also discuss the possible forms for the relativistic
wave functions which result from this analysis.

\smallskip\ 

\noindent We now follow Tsibidis' \cite{tsib} treatment of the Breit
equation, which itself is based on the work of Krolikowski \cite{krol}. The
spinor $\psi \left( \vec r\right) $ is a sixteen component wave function
which can be represented as a $4\times 4$ matrix 
\begin{equation}
\psi \left( \vec r\right) =\psi _{\gamma _0^{\left( 1\right) }\gamma
_0^{\left( 2\right) }}=\left( 
\begin{tabular}{l}
$\psi _{++}\psi _{+-}$ \\ 
$\psi _{-+}\psi _{--}$%
\end{tabular}
\right)  \label{mat}
\end{equation}
\noindent where the indices $+,-$ refer to the eigenvalues of $\gamma
_0^{(1)}$ and $\gamma _0^{\left( 2\right) }$. Equation (\ref{breit}) is
reduced to a set of 4 equations for $\psi _{++}$, $\psi _{+-}$,....
Following the technique of ref. [3], which has been already used to analyse
the non-relativistic case, we introduce the components 
\begin{eqnarray}
\left. 
\begin{array}{l}
\phi \\ 
\phi ^0
\end{array}
\right\} &=&P_0\frac i{\sqrt{2}}\left( \psi _{++}\mp \psi _{--}\right)
\label{new} \\
\left. 
\begin{array}{l}
\vec \phi \\ 
\vec \phi ^0
\end{array}
\right\} &=&\frac 12\left( \vec \sigma ^{(1)}-\vec \sigma ^{(2)}\right) P_1%
\frac 1{\sqrt{2}}\left( \psi _{+-}\pm \psi _{-+}\right)  \nonumber \\
\left. 
\begin{array}{l}
\chi \\ 
\chi ^0
\end{array}
\right\} &=&P_0\frac i{\sqrt{2}}\left( \psi _{+-}\mp \psi _{-+}\right) 
\nonumber \\
\left. 
\begin{array}{l}
\vec \chi \\ 
\vec \chi ^0
\end{array}
\right\} &=&\frac 12\left( \vec \sigma ^{\left( 1\right) }-\vec \sigma
^{\left( 2\right) }\right) P_1\frac 1{\sqrt{2}}\left( \psi _{++}\pm \psi
_{--}\right)  \nonumber
\end{eqnarray}
\noindent where $\phi $ and $\chi $ correspond to the spin zero states while 
$\vec \phi $ and $\vec \chi $ correspond to $S=1$ states (compare with eq. (%
\ref{decomp}). From Eqs. (\ref{breit}), (\ref{mat}) and (\ref{new}) we now
derive the following set of equations using the identities above for the
projection operators $P_0$ and $P_1$:

\begin{equation}
\begin{array}[t]{l}
\frac 12\left( E-S-P-V\right) \phi ^0-\frac{m_1+m_2}2\phi -i\vec p.\vec \phi
=0 \\ 
\\ 
\frac 12\left( E-S+P-V\right) \phi -\frac{m_1+m_2}2\phi ^0=0 \\ 
\\ 
\frac 12\left( E+S+P-V\right) \chi ^0-\frac{m_1-m_2}2\chi -i\vec p.\vec \chi
=0 \\ 
\\ 
\frac 12\left( E+S-P-V\right) \chi -\frac{m_1-m_2}2\chi ^0=0 \\ 
\\ 
\frac 12\left( E-S-P-V\right) \vec \chi -\frac{m_1+m_2}2\vec \chi ^0+i\vec p%
\chi ^0=0 \\ 
\\ 
\frac 12\left( E-S+P-V\right) \vec \chi ^0-\frac{m_1+m_2}2\vec \chi +i\vec p%
\times \vec \phi ^0=0 \\ 
\\ 
\frac 12\left( E+S+P-V\right) \vec \phi -\frac{m_1-m_2}2\vec \phi ^0+i\vec p%
\phi ^0=0 \\ 
\\ 
\frac 12\left( E+S-P-V\right) \vec \phi ^0-\frac{m_1-m_2}2\vec \phi +i\vec p%
\times \vec \chi ^0=0
\end{array}
\end{equation}
\noindent Each of the vector functions $\vec \phi ,\vec \chi ,\vec \phi ^0,%
\vec \chi ^0$, entering this equations, can be split into ``electric'',
``longitudinal'' and ``magnetic'' parts according to Eqs. (\ref{khi}) and (%
\ref{kr}) and expanded into spherical harmonics. We then obtain for each
value of the total angular momentum 12 radial wave functions for the vector
components : $\phi _e^j\left( r\right) ,\phi _L^j\left( r\right) ,\phi
_M^j\left( r\right) ,\chi _e^j\left( r\right) $ etc, together with the four
radial wave functions for the scalar components: $\phi ^{oj}\left( r\right)
,\phi ^j\left( r\right) ,\chi ^{oj}\left( r\right) ,\chi ^j\left( r\right) $%
. We write down the sixteen equations for these functions according to the
following classification.

\smallskip\ 

\noindent It is easy to see that $\phi ,\phi ^0,\phi _e,\phi _e^0,\phi
_L,\phi _L^0,\chi _M,\chi _M^0$ have parity P$\ =\eta \left( -1\right) ^j$
where $\eta $ is the intrinsic parity ($+1$ for two fermions and $-1$ for
fermion-antifermion system) and $j$ is the total momentum (we omit index $j$
at the radial components of the wave functions). the remaining 8 functions
namely $\chi ,\chi ^0,\chi _e,\chi _e^0,\chi _L,\chi _L^0,\phi _M,\phi _M^0,$
have P$=-\eta \left( -1\right) ^j$ and therefore the equations for the first
8 functions do not mix with the equations for the latter 8 components. The
former case may be called a pseudoscalar meson trajectory (PMT) and its
spectroscopic signature is $^1j_j$ or $^3j_j$ (the orbital momentum $\ell =j$%
) while the latter may be called a vector meson trajectory (VMT): it has the
spectroscopic signature $^3\left( j-1\right) _j$ or $^3\left( j+1\right) _j$ 
\cite{child}$.$ A fermion-antifermion system which conserves charge
conservation C will have 8 independent components so the 16 spinor
components will need to reduce to 8 dynamical equations and 8 constraint
equations, or 4 dynamical equations and 4 constraint equations for PMT and
the same again for VMT. We demonstrate this below.

\smallskip\ 

\noindent Thus we obtain after some algebra:

\newpage\ 

\underline{(i) PMT $^3j_j,$ P $=\eta \left( -1\right) $}$^j$

\begin{equation}
\begin{array}{l}
{\displaystyle {1 \over 2}}%
\left( E-S-P-V\right) \phi ^0-%
{\displaystyle {m_1+m_2 \over 2}}%
\phi -\left( 
{\displaystyle {d \over dr}}%
+%
{\displaystyle {2 \over r}}%
\right) \phi _e-%
{\displaystyle {1 \over r}}%
\phi _L=0 \\ 
\\ 
{\displaystyle {1 \over 2}}%
\left( E-S+P-V\right) \phi -%
{\displaystyle {m_1+m_2 \over 2}}%
\phi ^0=0 \\ 
\\ 
{\displaystyle {1 \over 2}}%
\left( E-S-P-V\right) \chi _M-%
{\displaystyle {m_2+m_2 \over 2}}%
\chi _M^0=0 \\ 
\\ 
{\displaystyle {1 \over 2}}%
\left( E-S+P-V\right) \chi _M^0-%
{\displaystyle {m_1+m_2 \over 2}}%
\chi _M+%
{\displaystyle {j\left( j+1\right)  \over r}}%
\phi _e^0+\left( 
{\displaystyle {d \over dr}}%
+%
{\displaystyle {1 \over r}}%
\right) \phi _L^0=0 \\ 
\\ 
{\displaystyle {1 \over 2}}%
\left( E+S+P-V\right) \phi _e-%
{\displaystyle {m_1-m_2 \over 2}}%
\phi _e^0+%
{\displaystyle {d\phi ^0 \over dr}}%
=0 \\ 
\\ 
{\displaystyle {1 \over 2}}%
\left( E+S+P-V\right) \phi _L-%
{\displaystyle {m_1-m_2 \over 2}}%
\phi _L^0-%
{\displaystyle {j\left( j+1\right)  \over r}}%
\phi ^0=0 \\ 
\\ 
{\displaystyle {1 \over 2}}%
\left( E+S-P-V\right) \phi _e^0-%
{\displaystyle {m_1-m_2 \over 2}}%
\phi _e+%
{\displaystyle {1 \over r}}%
\chi _M^0=0 \\ 
\\ 
{\displaystyle {1 \over 2}}%
\left( E+S-P-V\right) \phi _L^0-%
{\displaystyle {m_1-m_2 \over 2}}%
\phi _L-\left( 
{\displaystyle {d \over dr}}%
+%
{\displaystyle {1 \over r}}%
\right) \chi _M^0=0
\end{array}
\label{pmt}
\end{equation}
\underline{(ii) VMT $^3\left( j\pm 1\right) _j, $ P $=-\eta \left(
-1\right) $} $^j$

\begin{equation}
\begin{array}{l}
\\ 
{\displaystyle {1 \over 2}}%
\left( E+S+P-V\right) \chi ^{0}-%
{\displaystyle {m_{1}-m_{2} \over 2}}%
\chi -\left( 
{\displaystyle {d \over dr}}%
+%
{\displaystyle {2 \over r}}%
\right) \chi _{e}-%
{\displaystyle {1 \over r}}%
\chi _{L}=0 \\ 
\\ 
{\displaystyle {1 \over 2}}%
\left( E+S-P-V\right) \chi -%
{\displaystyle {m_{1}-m_{2} \over 2}}%
\chi ^{0}=0 \\ 
\\ 
{\displaystyle {1 \over 2}}%
\left( E-S-P-V\right) \chi _{e}-%
{\displaystyle {m_{1}+m_{2} \over 2}}%
\chi _{e}^{0}+%
{\displaystyle {d\chi ^{0} \over dr}}%
=0 \\ 
\\ 
{\displaystyle {1 \over 2}}%
\left( E-S-P-V\right) \chi _{L}-%
{\displaystyle {m_{1}+m_{2} \over 2}}%
\chi _{L}^{0}-%
{\displaystyle {j\left( j+1\right)  \over r}}%
\chi _{{}}^{0}=0 \\ 
\\ 
{\displaystyle {1 \over 2}}%
\left( E-S+P-V\right) \chi _{e}^{0}-%
{\displaystyle {m_{1}+m_{2} \over 2}}%
\chi _{e}+%
{\displaystyle {1 \over r}}%
\phi _{M}^{0}=0 \\ 
\\ 
{\displaystyle {1 \over 2}}%
\left( E-S+P-V\right) \chi _{L}^{0}-%
{\displaystyle {m_{1}+m_{2} \over 2}}%
\chi _{L}-\left( 
{\displaystyle {d \over dr}}%
+%
{\displaystyle {1 \over r}}%
\right) \phi _{M}^{0}=0 \\ 
\\ 
{\displaystyle {1 \over 2}}%
\left( E+S-P-V\right) \phi _{M}^{0}-%
{\displaystyle {m_{1}-m_{2} \over 2}}%
\phi _{M}^{{}}+%
{\displaystyle {j(j+1) \over r}}%
\chi _{e}^{0}-\left( 
{\displaystyle {d \over dr}}%
+%
{\displaystyle {1 \over r}}%
\right) \chi _{L}^{0}=0 \\ 
\\ 
{\displaystyle {1 \over 2}}%
\left( E+S+P-V\right) \phi _{M}^{{}}-%
{\displaystyle {m_{1}-m_{2} \over 2}}%
\phi _{M}^{0}=0
\end{array}
\label{vmt}
\end{equation}

\noindent The non-relativistic analysis of the previous section for
pseudoscalar exchange $P\neq 0$ refers to the spin $S=1$ states connected to
the ''large'' components of the relativistic wave function $\psi _{++}$.
This implies that $\chi _{e}$, $\chi _{L}$ of Eq. (\ref{khi3}) are the
non-relativistic analogues of the same functions entering Eq. (\ref{vmt}).
Remember also that the $1/r^{3}$ singularity of Eq. (\ref{khi3}) only
involves the functions $\chi _{e}$ and $\chi _{L}$, so it is sufficient for
our purposes to concentrate on the following set of equations for the case $%
m_{1}=m_{2}=m$:

\begin{equation}
\begin{array}{l}
{\displaystyle {1 \over 2}}%
\left( E+S+P-V\right) \chi _{}^0-\left( 
{\displaystyle {d \over dr}}%
+%
{\displaystyle {2 \over r}}%
\right) \chi _e^{}-%
{\displaystyle {1 \over r}}%
\chi _L=0 \\ 
\\ 
{\displaystyle {1 \over 2}}%
\left( E-S-P-V\right) \chi _e-m\chi _e^0+%
{\displaystyle {d\chi ^0 \over dr}}%
=0 \\ 
\\ 
{\displaystyle {1 \over 2}}%
\left( E-S-P-V\right) \chi _L-m\chi _L^0-%
{\displaystyle {j\left( j+1\right)  \over r}}%
\chi ^0=0 \\ 
\\ 
{\displaystyle {1 \over 2}}%
\left( E-S+P-V\right) \chi _e^0-m\chi _e+%
{\displaystyle {1 \over r}}%
\phi _M^0=0 \\ 
\\ 
{\displaystyle {1 \over 2}}%
\left( E-S+P-V\right) \chi _L^0-m\chi _L-\left( 
{\displaystyle {d \over dr}}%
+%
{\displaystyle {1 \over r}}%
\right) \phi _M^0=0 \\ 
\\ 
{\displaystyle {1 \over 2}}%
\left( E+S-P-V\right) \phi _M^0+%
{\displaystyle {j\left( j+1\right)  \over r}}%
\chi _e^0+\left( 
{\displaystyle {d \over dr}}%
+%
{\displaystyle {1 \over r}}%
\right) \chi _L^0=0
\end{array}
\label{eqm}
\end{equation}

\noindent \ First of all we should try to show that the non-relativistic
limit of (\ref{eqm}) gives Eq. (\ref{khi3}) for $\chi _e$and $\chi _L$. This
is not a trivial exercise but we can proceed as follows. Add and subtract
the second and the fourth equation of (\ref{vmt}) and introduce $\chi
_e^{\pm }=\frac 12\left( \chi _e\pm \chi _e^0\right) $. Then we can get the
equation (keeping only $P\neq 0$): 
\begin{equation}
\left( E-2m\right) \chi _e^{+}+\frac{d\chi ^0}{dr}\left( 1+\frac P{E+2m}%
\right) +\frac 1r\phi _M^0\left( 1-\frac P{E+2m}\right) -\frac{P^2}{E+2m}%
\chi _e^{+}=0  \label{re1}
\end{equation}
\noindent From (\ref{eqm}) and (\ref{re1}) we can obtain for $\chi _L^{+}=%
\frac 12\left( \chi _L+\chi _L^0\right) $:

\begin{equation}
\left( E-2m\right) \chi _L^{+}-\frac{j(j+1)}r\left( 1+\frac P{E+2m}\right)
\chi ^0-(\frac d{dr}+\frac 1r)\phi _M^0\left( 1-\frac P{E+2m}\right) -\frac{%
P^2}{E+2m}\chi _L^{+}=0  \label{re2}
\end{equation}

\noindent \ In the nonrelativistic approximation $\chi ^0$and $\phi _M^0$
are readily expressed through $\chi _e^{+}$and $\chi _L^{+}$ (this is
correct only for the non-relativistic limit). We get: 
\begin{eqnarray}
\chi ^0 &\simeq &\frac 1m\left[ \left( \frac{d\chi _e^{+}}{dr}+\frac 2r\chi
_e^{+}+\frac 1r\chi _L^{+}\right) \left( 1-\frac P{4m}\right) +\frac{%
P^{\prime }}{4m}\chi _e^{+}\right]  \label{rel2} \\
\phi _M^0 &\simeq &\frac 1m\left[ \left( \frac{j\left( j+1\right) }r\chi
_e^{+}+\left( \frac d{dr}+\frac 1r\right) \chi _L^{+}\right) \left( 1+\frac P%
{4m}\right) -\frac{P^{\prime }}{4m}\chi _L^{+}\right]  \nonumber
\end{eqnarray}
\noindent where $P^{\prime }=\frac{dP}{dr}$. In these equations we neglect
all non-relativistic corrections and keep only linear terms in $P$. Then
substituting (\ref{rel2}) into (\ref{re1}) and (\ref{re2}) we see that only
derivatives of $P$ remain in the final equations for $\chi _e^{+}$and $\chi
_L^{+}$. Eventually we obtain the non-relativistic limit: 
\begin{equation}
\begin{array}{l}
m\left( E-2m\right) \chi _e^{+}+%
{\displaystyle {d^2\chi _e^{+} \over dr^2}}%
+%
{\displaystyle {2 \over r}}%
{\displaystyle {d\chi _e^{+} \over dr}}%
-%
{\displaystyle {j\left( j+1\right) +2 \over r^2}}%
\chi _e^{+}-%
{\displaystyle {2 \over r^2}}%
\chi _L^{+}+\left( 
{\displaystyle {P^{\prime \prime } \over 4m}}%
-%
{\displaystyle {P^{\prime } \over 8mr}}%
\right) \chi _e^{+}=0, \\ 
m\left( E-2m\right) \chi _L^{+}+%
{\displaystyle {d^2\chi _L^{+} \over dr^2}}%
+%
{\displaystyle {2 \over r}}%
{\displaystyle {d\chi _L^{+} \over dr}}%
-%
{\displaystyle {j\left( j+1\right)  \over r^2}}%
\chi _L^{+}-%
{\displaystyle {2j\left( j+1\right)  \over r^2}}%
\chi _e^{+}-%
{\displaystyle {P^{\prime \prime } \over 4m}}%
\chi _L^{+}=0
\end{array}
\label{re3}
\end{equation}

\noindent For the exchange of a pseudoscalar particle: 
\begin{equation}
P=\frac{m^2}{\pi v_A^2}\frac 1re^{-m_Ar}
\end{equation}
\noindent So we see that the terms proportional to $P^{\prime }/r$ and $%
P^{\prime \prime }$in Eq. (\ref{re3}) give the $1/r^3$ singularity of eq. (%
\ref{khi3}). We show in the Appendix that the relativistic equations of Eq. (%
\ref{vmt}) do not give a $1/r^3$ singularity. Thus this singularity is just
an artefact of the non-relativistic approximation.

\section{Conclusions}

\noindent We conclude that heavy fermions $f$ of mass larger than 800 GeV
are required if they are to bind with $\bar f$ by Higgs or pseudoscalar
exchange. If fermions of this mass exist, a whole new region of physics will
be associated with this high mass scale: the fermions themselves and their
decay modes, and the $f\,\bar f$ bound states, their decays and systematics.
The details of this new physics, which involves binding by an unfamiliar
mechanism, cannot be predicted in advance.

\smallskip\ 

\noindent We have also demonstrated that the $1/r^3$ singularity seemingly
present in pseudoscalar exchange will vanish if a relativistic framework is
adopted for the calculation. Pseudoscalar exchange in relativistic quantum
field theory arises from renormalisable theories and hence cannot lead to
collapse.\ 

\smallskip\ 

\noindent We would like to thank G. Tsibidis for his help with the
Breit-Krolikowski equations. This work has been supported in part by INTAS
grant 93-283 and INTAS-REBR 943986. One of us (AA) thanks PPARC and the
University of Sussex for support during his stay at Sussex in 1997.

\section{Appendix}

\noindent We now look at the relativistic case to show that there is no $%
1/r^3$ singularity at the origin arising from pseudoscalar exchange. It is
not easy to derive relativistic equations for the two functions $\chi _e$and 
$\chi _L$ of the non-relativistic approximation (\ref{re3}). It is easier to
exclude $\chi _e$, $\chi _e^0$, $\chi _L$, $\chi _L^0$ from Eq. (\ref{eqm})
and then obtain a pair of equations for $\chi ^0$, $\phi _M^0$.

\smallskip\ 

\noindent To this end we have from the second and fourth equations of (\ref
{eqm}) (keeping $P\neq 0$ and $S\neq 0)$: 
\begin{equation}
\begin{array}{l}
\chi _e=%
{\displaystyle {4m \over 4m^2-\left( E-S\right) ^2+P^2}}%
\left[ 
{\displaystyle {1 \over 2m}}%
\left( E-S+P\right) 
{\displaystyle {d\chi ^0 \over dr}}%
+%
{\displaystyle {\phi _M^0 \over r}}%
\right] \\ 
\chi _e^0=%
{\displaystyle {4m \over 4m^2-\left( E-S\right) ^2+P^2}}%
\left[ 
{\displaystyle {1 \over 2m}}%
\left( E-S-P\right) 
{\displaystyle {\phi _M^0 \over r}}%
+%
{\displaystyle {d\chi ^0 \over dr}}%
\right]
\end{array}
\end{equation}
and from the third and the fifth equations: 
\begin{eqnarray}
\chi _L &=&\frac{-4m}{4m_{}^2-\left( E-S\right) ^2+P^2}\left[ \frac 1{2m}%
\left( E-S+P\right) \frac{j\left( j+1\right) }r\chi ^0+\left( \frac d{dr}+%
\frac 1r\right) \phi _M^0\right] \\
\chi _L^0 &=&\frac{-4m}{4m_{}^2-\left( E-S\right) _{}^2+P^2}\left[ \frac 1{2m%
}\left( E-S-P\right) \left( \frac d{dr}+\frac 1r\right) \phi _M^0+\frac{%
j\left( j+1\right) }r\chi ^0\right]  \nonumber
\end{eqnarray}
Substituting these expressions into the first and the last of equations (\ref
{eqm}) we get two second order equations for $\chi ^0$and $\phi _M^0$: 
\begin{equation}
\begin{array}{l}
{\displaystyle {d^2\chi ^0 \over dr^2}}%
+%
{\displaystyle {2 \over r}}%
{\displaystyle {d\chi ^0 \over dr}}%
-%
{\displaystyle {j(j+1) \over r^2}}%
\chi ^0-%
{\displaystyle {1 \over 4}}%
\left[ 4m^2-\left( E-S\right) ^2+P^2\right] 
{\displaystyle {E+S+P \over E-S+P}}%
\chi ^0 \\ 
-%
{\displaystyle {d \over dr}}%
\left[ \ln 
{\displaystyle {4m^2-\left( E-S\right) ^2+P^2 \over E-S+P}}%
\right] 
{\displaystyle {d\chi ^0 \over dr}}%
- \\ 
-%
{\displaystyle {4m \over E-S+P}}%
{\displaystyle {d \over dr}}%
\left[ \ln \left( 4m^2-\left( E-S\right) ^2+P^2\right) \right] 
{\displaystyle {\phi _M^0 \over r}}%
=0 \\ 
\\ 
{\displaystyle {d^2\phi _M^0 \over dr^2}}%
+%
{\displaystyle {2 \over r}}%
{\displaystyle {d\phi _M^0 \over dr}}%
-%
{\displaystyle {j\left( j+1\right)  \over r^2}}%
\phi _M^0-%
{\displaystyle {1 \over 4}}%
\left[ 4m^2-\left( E-S\right) ^2+P^2\right] 
{\displaystyle {E+S-P \over E-S-P}}%
\phi _M^0- \\ 
-%
{\displaystyle {d \over dr}}%
\left[ \ln 
{\displaystyle {4m^2-\left( E-S\right) ^2+P^2 \over E-S-P}}%
\right] \left( 
{\displaystyle {d \over dr}}%
+%
{\displaystyle {1 \over r}}%
\right) \phi _M^0-%
{\displaystyle {4m \over E-S-P}}%
{\displaystyle {d \over dr}}%
\\ 
\left[ \ln \left( 4m^2-\left( E-S\right) ^2+P^2\right) \right] 
{\displaystyle {j(j+1) \over r^{}}}%
\chi ^0=0
\end{array}
\label{last}
\end{equation}
\noindent It is now clear what a delicate problem it is to go to the
nonrelativistic limit for the pseudoscalar case. For the scalar interaction, 
$S$ should be retained only in $4m^2-\left( E-S\right) ^2\simeq -4m\left(
\epsilon -S\right) $ where $\epsilon =E-2m$ in the first line of Eq. (\ref
{last}). This results in the usual Schr\"odinger equation with potential
energy $S$. For $S=0,P\neq 0$ we should keep the apparently small term $P/2m$
even in the limit $m\rightarrow \infty $ term since the coupling constant in 
$P$ is proportional to $m^2$. We do not however keep $P^2/4m^{2\text{ }}$%
terms in order to get the nonrelativistic equations (\ref{khi4}). The most
important thing which we learn from (\ref{last}) is that when $r\rightarrow
0 $ there are only singularities $\sim 1/r^2$, and therefore there is no
collapse. A more detailed analysis of the relativistic equation is outside
the scope of this paper: Further details can be found in the papers by
Mourad and Sazdjian \cite{mour} and Crater and Long \cite{xxx}.

\end{document}